\documentstyle[twocolumn,aps,epsf]{revtex}

\begin{document}
\twocolumn[\hsize\textwidth\columnwidth\hsize\csname
@twocolumnfalse\endcsname
%
%
%============================================================
%  Title 
%============================================================
\title{Effect of Coulomb blockade on STM current through a granular
  film} \author{H. Imamura$^{1,2}$, J. Chiba$^{2}$, S. Mitani$^{2}$,
  K. Takanashi$^{2}$, S. Takahashi$^{2}$ , S. Maekawa$^{2}$,
  H. Fujimori$^{2}$} \address{$^{1}$CREST, $^{2}$Institute for Materials
  Research, Tohoku University, Sendai 980-8577, Japan} \maketitle
%
%
%============================================================
%  Abstract
%============================================================
\begin{abstract}
 The electron transport through an array of tunnel junctions consisting of an
 STM tip and a granular film is studied both theoretically and
 experimentally.  When the tunnel resistance between the tip and a
 granule on the surface is much larger than those between granules, a
 bottleneck of the tunneling current is created in the array. It is
 shown that the period of the Coulomb staircase(CS) is given by the
 capacitance at the bottleneck.
 Our STM experiments on Co-Al-O granular films show the CS with a
 single period at room temperature.  This provides a new possibility for
 single-electron-spin-electronic devices at room temperature.
\end{abstract} 
\vskip2pc]
%
%
%============================================================
%  Introduction
%============================================================

Charging effects on single electron tunneling such as Coulomb blockade
and Coulomb oscillation have attracted much
interest\cite{sct_naz}. Recent advances in nano-technology
enable us to fabricate small tunnel junctions where charging
effects play an essential role.
%A highly-resistive granular film is one of such
%single electron tunneling systems\cite{abeles}.  
The I-V characteristics for double tunnel junction systems have been
extensively studied, where the step-like structure called Coulomb
staircase(CS) is observed when the resistances between the junctions are
not equal\cite{sct_naz}.
For these asymmetric double junction systems, the junction with a
large resistance behaves like a bottleneck of the tunneling
current and the central island is charged through the other junction
up to the maximum charge.  The tunneling current jumps 
when the maximum charge changes.
Recently, the CS has been observed by using a tip of scanning tunneling
microscope(STM) of nanometer-size in highly resistive
granular films\cite{bar-sadeh94,bar-sadeh95,desmicht98,chiba98} as
well as metal-droplet systems \cite{wilkins89,amman93}.
%\cite{wilkins89,amman93,dubois93}.
For a granular film, which is considered to be an array of tunnel
junctions,
the observed CS implies that a bottleneck exists in the
conducting paths. However, the physics behind the CS in a granular
film is not clear because it contains many granules with
different size and the conducting path may form a three-dimensional
network in a thick granular film.

Bar-Sadeh {\it et al}. have studied the STM current through a
nonmagnetic granular film, Au-Al$_{2}$O$_{3}$, by using the cryogenic
STM\cite{bar-sadeh94,bar-sadeh95}.  They observed the CS
%structure in
%the I-V curve 
at temperatures $T=4.2$ and 78 K, and analyzed the
experimental data by using a triple barrier model.  They assumed that
the rate for tunneling between two granules is small and the number of excess 
electrons in
each granule is treated independently.  Because of these assumptions, the
CS was given by the superposition of two different periods in their
model: one was determined by the tunnel process between the STM tip and a
granule, the other between another granule and the base electrode.
On the contrary, as we will show later, the CS has a single period which 
is determined by the capacitance at the bottleneck.

%
%
%============================================================
%  In this Letter
%============================================================
In this Letter, we study the electron transport through an array of
tunnel junctions consisting of an STM tip and a granular film both
theoretically and experimentally.  In this
system, we can vary the tunnel resistance between the
tip and a granule on the surface by changing the distance between them. 
When the tunnel resistance between the tip and a granule on the surface
is much larger than those between granules, a bottleneck of the
tunneling current is created in the array.
Theoretically, we find that the period of the CS is given by
the capacitance of the bottleneck even in a
thick film with many granules between the tip and the base electrode.
We present results from STM experiments on 10 nm- and 1$\mu$m-thick
Co-Al-O granular films which have the CS
with a single period at room temperature.
We propose that tunnel
magnetoresistance (TMR) oscillates with the same period as the CS for
magnetic granular films.  Our theoretical and experimental studies
provide a new direction for single-electron-spin-electronic devices at
room temperature.

%
%
%
%
%============================================================
%  Our setup
%============================================================

%
%
%%%%%%%%%%%%%%%%%%%%%%%%%%%%%%%%%%%%%%%%%
%%  figure 1 : schematics              %%
%%%%%%%%%%%%%%%%%%%%%%%%%%%%%%%%%%%%%%%%%
\begin{figure}
    \epsfxsize=0.9\columnwidth \centerline{\hbox{
\epsffile{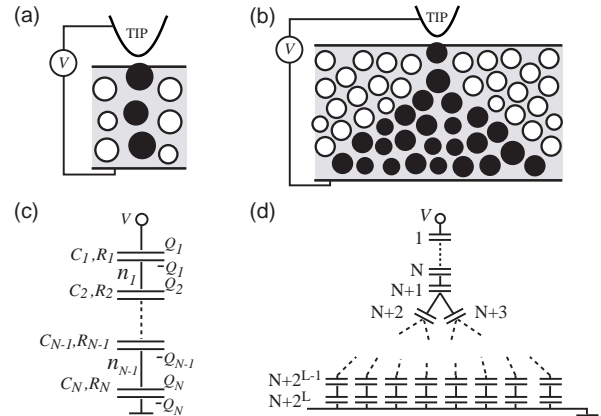}} } \caption{ The system with an STM tip and a
thin(thick) granular film is schematically shown in panel (a)((b)).
Filled(hollow) circles represent granules in(out of) the conducting
path.  The insulating matrix is indicated by the shading.  The corresponding
theoretical models for (a) and (b) are shown in panels (c) and (d),
respectively.  For the thick film, a Bethe-lattice network of granules,
where each granule has three neighbors, is connected to the
one-dimensional array.}  \label{fig:cartoons}
\end{figure}
Our setup is schematically shown in Figs. \ref{fig:cartoons}(a) and
\ref{fig:cartoons}(b). 
The current flows from the STM tip to the base electrode through a
granular film.  
The system with a thin granular film in panel (a) is modeled by the one-dimensional array of tunnel junctions as in
panel (c).  
%~~~~~~~~~~~~~~~~~~~~~~~~~~~~~~~~~~~~~~~~
%~~~~~~~~~~~~~~~~~~~~~~~~~~~~~~~~~~~~~~~~
We will show that our experimental results for the 10 nm-thick film are well
explained by this model with $N=3$.
%~~~~~~~~~~~~~~~~~~~~~~~~~~~~~~~~~~~~~~~~
%~~~~~~~~~~~~~~~~~~~~~~~~~~~~~~~~~~~~~~~~
On the other hand, such a one-dimensional array is not
appropriate for a thick granular film, because,
as illustrated in Fig. \ref{fig:cartoons}(b), the conducting paths spread
and form a three-dimensional network as the distance from the tip
increases.  We model this system by a one-dimensional array of
$N$-junctions connected to a Bethe-lattice network with 3 nearest
neighbors
as shown in Fig. \ref{fig:cartoons} (d).
Each junction is
characterized by a tunnel resistance $R_{j}$, capacitance $C_{j}$, and
carrying charge $Q_{j}$.  The number of excess electrons in the $k$-th
granule is represented by $n_{k}$.

%
%
%============================================================
%  Free energy
%============================================================

The free energy for the state characterized by the set of charges
$\{n_{i}\}\equiv (n_{1},n_{2},\dots)$ is given by
\begin{equation}
F(\{n_{i}\}) =\sum_{i}\frac{Q_{i}^{2}}{2C_{i}} - (Q_{1}- e \xi)V,
  \label{eq:free}
\end{equation}
where $Q_{1}$ represents the charge at the surface, $\xi$ is the number
of electrons supplied by the
voltage source and $i$ goes from 1 to $N$.
When an electron tunnels through the
$k$-th junction, the charge $Q_{i}$ deviates from its initial value by
$\delta Q_{i}^{k}$.  Let us consider the energy change due to the
single electron tunneling, $E_{k}^{\pm}$, where the superscript $+(-)$
denotes the process that an electron tunnels upward(downward)
through the $k$-th junction
in Figs. \ref{fig:cartoons}(c) and \ref{fig:cartoons}(d).  
From Eq. (\ref{eq:free}), we obtain
\begin{eqnarray}
E_{k}^{\pm}(\{n_{i}\})
&=&\sum_{i}\left(\widetilde{\sum_{j<i}}\frac{\delta
Q_{j}^{k}}{C_{j}}\right) n_{i} +\frac{1}{2}\sum_{i}\frac{\left(\delta
Q_{i}^{k}\right)^{2}}{C_{i}}\nonumber\\ &&-(\delta Q_{1}^{k}\pm
e\delta_{1,k})V, \label{eq:epm}
\end{eqnarray}
where $\widetilde{\sum_{i}}$ represents the summation along the
conducting path, and $\delta_{1,k}$ is Kronecker's delta function. 
The deviation $\delta Q_{i}^{k}$ is determined by 
Kirchhoff's law and is independent of the number of excess electrons
$\{n_{i}\}$.
The tunneling rate is obtained by using the
golden rule \cite{sct_naz} as
\begin{equation}
\Gamma_{k}^{\pm}(\{n_{i}\})\nonumber\\
=\frac{E_{k}^{\pm}(\{n_{i}\})}{e^{2}R_{k}
[\exp(E_{k}^{\pm}(\{n_{i}\})/T)-1]}.
\end{equation}
By solving the master equation for the probability of states
$p(\{n_{i}\})$ \cite{sct_naz}, the tunneling current through the $k$-th
junction is obtained as
\begin{equation}
I_{k}=e\sum_{\{n_{i}\}}p(\{n_{i}\}) \left[ \Gamma_{k}^{+}(\{n_{i}\}) -
 \Gamma_{k}^{-}(\{n_{i}\}) \right]\label{eq:crnt}.
\end{equation}
For simplicity, we neglect the effect of residual fractional
charge\cite{bar-sadeh94,bar-sadeh95,hanna91},
cotunneling\cite{takahashi98,mitani98,averin90}, spin
accumulation\cite{johnson95,hima_tmr,takahashi99} and level
quantization\cite{kubo62} in the granules.

%
%
%============================================================
%  Parameters
%============================================================

\begin{figure}
    \epsfxsize=0.8\columnwidth \centerline{\hbox{
\epsffile{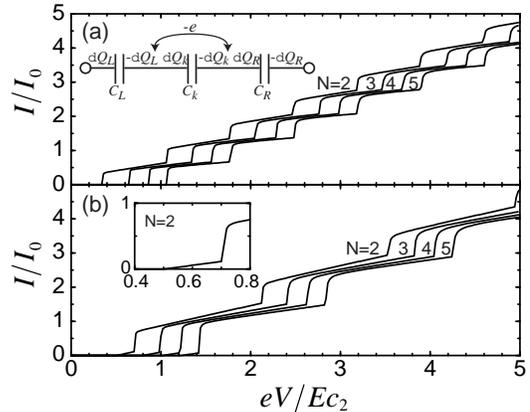}} } \caption{ The tunneling current normalized
with $I_{0}=E_{C_{2}}/eR_{1}$, where $E_{C_{2}}=e^{2}/C_{2}$,  is
 plotted against the bias voltage at $T=0$.
The number of junctions are $N=2\sim 5$ from left to right.  
%%All systems have a bottleneck at the first junction.  
The tunnel resistance
ratio is taken to be $R_{1}:R_{2}:R_{3}:R_{4}:R_{5}=1000:1:1:1:1$.  The
 capacitance ratio is (a)
$C_{1}:C_{2}:C_{3}:C_{4}:C_{5}=\protect\sqrt{2}:1:\protect\sqrt{3}:\protect\sqrt{5}:\protect\sqrt{6}$,
(b)
$C_{1}:C_{2}:C_{3}:C_{4}:C_{5}=1/\protect\sqrt{2}:1:\protect\sqrt{3}:\protect\sqrt{5}:\protect\sqrt{6}$.
The tunneling current with $n_{1}=0$ as a dominant  state for $N=2$ is
shown in the inset of panel (b) } \label{fig:crnt2-5}
\end{figure}

We first look at the I-V characteristics for a one-dimensional
array of tunnel junctions with
a bottleneck of the tunneling current between the tip and a granule on the 
surface.
%
%
%============================================================
% current classification
%============================================================
In Figs. \ref{fig:crnt2-5}(a) and \ref{fig:crnt2-5}(b), we present the
numerical results for
tunnel junctions with $N=2\sim 5$ at $T=0$, $N$ being the number of
junctions.  Due to the
charging energy in each junction, single electron tunneling is blocked
as long as the bias voltage $V$ is lower than the threshold value
$V_{T}$\cite{melsen97}.  From Eq. (\ref{eq:epm}) the bias voltage
$V_{T}^{k}$, above which 
the initial  state $(0,\dots,0)$ is unstable for
electron tunneling through $k$-th junction, i.e., 
$E_{k}^{+}(0,\dots,0) < 0$, is given by
\begin{equation}
V_{T}^{k}=\frac{\sum_{i}(\delta Q_{i}^{k})^{2}/C_{i}}{2(\delta
Q_{1}^{k}+e\delta_{1,k})}
=\frac{e}{2}\left(\frac{1}{C_{T}}-\frac{1}{C_{k}}\right).  \label{eq:vtk}
\end{equation}
The last expression in Eq. (\ref{eq:vtk}) is easily obtained by 
considering the equivalent network shown in the inset of
Fig. \ref{fig:crnt2-5}(a), where $1/C_{L}=\sum_{i<k}1/C_{i}$ and
$1/C_{R}=\sum_{i>k}1/C_{i}$.  
%Then, the numerator of Eq. (\ref{eq:vtk}) is
%written as $(\delta Q_{L})^{2}/C_{L} + (\delta Q_{k})^{2}/C_{k} +
%(\delta Q_{R})^{2}/C_{R}$, and we have $V_{T}^{k}=(e/2)(1/C_{T}
%-1/C_{k})$.  
%Therefore, the threshold voltage is obtained as
%$V_{T}= \min_{k} V_{T}^{k}$.  
%We have $V_{T}^{k}=(e/2)(1/C_{T}-1/C_{k})$ and the threshold voltage is
%obtained as $V_{T}= \min_{k} V_{T}^{k}$.  
The threshold voltage is given by $V_{T}= \min_{k} V_{T}^{k}$.  
As pointed out by Melsen {\it et al.}, the Coulomb blockade
region increases as the number of junctions, $N$, increases\cite{melsen97}
as shown in Figs. \ref{fig:crnt2-5}(a) and \ref{fig:crnt2-5}(b).  
For large $N$, the
threshold voltage is expressed in terms of the average capacitance
$\langle C \rangle$ as $V_{T} \simeq (e/2\langle C \rangle)(N-1)$ and is 
proportional to the film thickness, because the total
capacitance is inversely proportional to the
layer thickness.

For a thick film with many granules between the tip and the
base electrode, the conducting paths form a three-dimensional
network as shown in Figs. \ref{fig:cartoons}(b) and \ref{fig:cartoons}(d).  The
decrease of the total capacitance for a junction network with increasing 
film thickness is much weaker than that for a one-dimensional array.
Therefore, the threshold voltage $V_{T}$ of a thick film remains of the
same order of the magnitude as that for a thin film.
Later we will show that
the experimental results for 1 $\mu$m-thick film are well explained by considering the junction network as the Bethe-lattice as shown
in Fig. \ref{fig:cartoons}(d).

%
%
%%%%%%%%%%%%%%%%%%%%%%%%%%%%%%%%%%%%%%%%
%%  figure 2 : general I-V            %%
%%%%%%%%%%%%%%%%%%%%%%%%%%%%%%%%%%%%%%%%

The CS is classified into two types as shown in
Figs. \ref{fig:crnt2-5}(a) and \ref{fig:crnt2-5}(b). 
The criterion is whether the
capacitance $C_{1}$ of the bottleneck is the smallest of all of the
junctions, i.e., whether $V_{T}=V_{T}^{1}$ or $V_{T}=\min_{k\neq
1}V_{T}^{k}$.
A typical CS for $V_{T}=\min_{k\neq 1}V_{T}^{k}$ is
given in Fig. \ref{fig:crnt2-5}(a), where electrons start to
accumulate at the bottleneck once the bias voltage exceeds $V_{T}$.
Until the accumulated electrons tunnel out through the bottleneck, the
voltage drop caused by them forbids electrons to tunnel through the
other junctions.  Therefore, the stable state is given by
$(n_{1},0,\dots,0)$ and the tunneling current jumps at the bias voltage
where the number of accumulated electrons $n_{1}$ changes.  This
number $n_{1}$ is the minimum value satisfying
the conditions
\begin{equation}
E_{1}^{+}(n_{1},0,\dots,0) 
< 0,\ \ E_{k\neq 1}^{+}(n_{1},0,\dots,0)\ge 0.  \label{eq:cond_n1}
\end{equation}
The first condition represents the
accumulated electrons tunneling out through the first junction.  The
second indicates that electrons cannot tunnel through the other
junctions, and is rewritten as $ (e^{2}/C_{1})n_{1} \ge eV -
V_{T}^{k\neq 1}$.   Therefore, the tunneling current jumps at the bias
voltage $V$ given by
\begin{equation}
V=\min_{k\neq 1}V_{T}^{k} + (e^{2}/C_{1})(n_{1} -1).  \label{eq:period}
\end{equation}
From Eq. (\ref{eq:period}), one can easily see that the CS has a single
period of $e/C_{1}$.
The tunneling current for each plateau of the CS is approximately given
by $I\simeq e\Gamma_{1}^{+}(n_{1},0,\dots,0)$.

%============================================================
%  current classification
%============================================================

On the other hand, if 
$C_{1}$ is the smallest, the
tunneling current does
not jump at $V_{T}$ as shown in Fig. \ref{fig:crnt2-5}(b).  For
$V_{T}<V<\min_{k\neq 1}V_{T}^{k}$, the tunneling current is approximately
given by $I\simeq e\Gamma_{1}^{+}(0,\dots,0)$.  Once $V$
exceeds $\min_{k\neq 1}V_{T}^{k}$, electrons start to accumulate at the
bottleneck and the I-V curve shows a CS. The bias voltage $V$ 
at which the tunneling current jumps is given by
Eq.(\ref{eq:period})\cite{double}.

%~~~~~~~~~~~~~~~~~~~~~~~~~~~~~~~~~~~~~~~~~~~~~~~~~~~~~~~~~~~~
%~~~~~~~~~~~~~~~~~~~~~~~~~~~~~~~~~~~~~~~~~~~~~~~~~~~~~~~~~~~~
%
%============================================================
%  Hanna
%============================================================
%For a double junction system $(N=2)$, the shape of the CS has been
%studied by Hanna and Tinkham\cite{hanna91}.  Neglecting residual
%fractional charge, they classified the CS into two types depending on
%whether the capacitance ratio $C_{1}/C_{2}$ is larger than one.  This is
%consistent with our classification for $N=2$.
%============================================================
%  Shape of Coulomb staircase  (bottleneck at other junction)
%============================================================
When the bottleneck is placed at another junction $j
(1<j<N)$, the stable state is
$(0,\dots,0,n_{j-1}=-n_{j},n_{j},0,\dots,0)$ and the period of the CS is
determined by the capacitance $C_{j}$ at the bottleneck.
The CS is classified into two types in the same way as those shown in
Figs. \ref{fig:crnt2-5}(a) and  \ref{fig:crnt2-5}(b).
However, the criterion is whether the capacitance $C_{j}$ is the
smallest or not.

%
%
%
%============================================================
%  Thick
%============================================================
The period of CS in a thick film shown in Figs. \ref{fig:cartoons}(b) and
\ref{fig:cartoons}(d) is also 
determined  from Eqs. (\ref{eq:epm}) and (\ref{eq:cond_n1}),
as long as the tip is coupled to a 
single granule on the surface and the bottleneck is created between
them.  Therefore, the voltage where the tunneling current jumps is given 
by Eq. (\ref{eq:period}).  The CS has a single period determined by the
capacitance at the bottleneck, even for a thick film.

%~~~~~~~~~~~~~~~~~~~~~~~~~~~~~~~~~~~~~~~~~~~~~~~~~~~~~~~~~~~~
%~~~~~~~~~~~~~~~~~~~~~~~~~~~~~~~~~~~~~~~~~~~~~~~~~~~~~~~~~~~~
%
%
%
%============================================================
%  10 nm Co-Al-O film (Experiment)
%============================================================

We have performed STM experiments on Co-Al-O granular films.
Samples with
different thicknesses were prepared by an oxygen-reactive sputtering with
a Co-Al alloy target; 10 nm- and 1 $\mu$m-thick
Co$_{36}$Al$_{22}$O$_{42}$ films consisting of Co granules embedded in
an Al-oxide matrix were deposited on glass substrates. For the 10 nm-thick
Co$_{36}$Al$_{22}$O$_{42}$ film, a 200 nm-thick Co-Al alloy layer was
inserted between the Co-Al-O granular film and the glass substrate as a
base electrode. A conventional STM system was used for I-V
measurements under high vacuum. The I-V curves were obtained by using a
platinum tip at room temperature, and by placing the
tip on a Co granule.
The tunnel resistance between granules for Co-Al-O granular films is
estimated to be about $10^{5} \sim 10^{6} \Omega$  from the average diameter
of granules ($\sim$ 3 nm), the average intergranular distance ($\sim$
1 nm), and electrical resistivity \cite{chiba98,mitani98}. 
%$$ \cite{chiba98,mitani98,ohnuma97}. 
On the other
hand, the tunnel resistance between the tip and a granule on the surface,
$R_{1}$, is about $10^{8} \sim 10^{9} \Omega$.
Therefore, $R_{1}$ is
$10^{2} \sim 10^{4}$ times larger than the other tunnel resistances.

%
%
%
%============================================================
%  10 nm Co-Al-O film (Theory)
%============================================================

\begin{figure}
    \epsfxsize=0.8\columnwidth \centerline{\hbox{
\epsffile{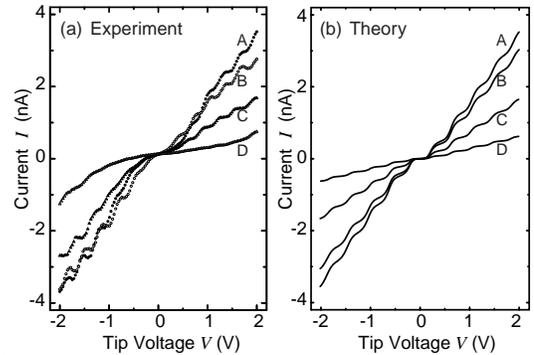}} } \caption{ (a)Experimental I-V curves for
a 10 nm-thick Co$_{36}$Al$_{22}$O$_{42}$ at room temperature.
A, B, C and D refer to different distance between the STM tip from the
surface of the sample.
The lateral position for A and B is different from that for 
C and D.  (b)
Corresponding theoretical curves in a triple tunnel junction system at
$T=300$ K.  The tunnel resistance at the bottleneck is taken to be
 $R_{1}=600, 700, 1300$, and 3500 M$\Omega$ for the line A, B, C and D,
respectively.  The other tunnel resistances are $R_{2}=R_{3}=1{\rm
M}\Omega$ and the capacitances are $C_{1}=4.48\times 10^{-19}$ F,
$C_{2}=2.13\times 10^{-19}$ F and $C_{3}=3.62\times 10^{-19}$ F for all
curves.}  \label{fig:thin}
\end{figure}

\begin{figure}
    \epsfxsize=0.8\columnwidth \centerline{\hbox{
\epsffile{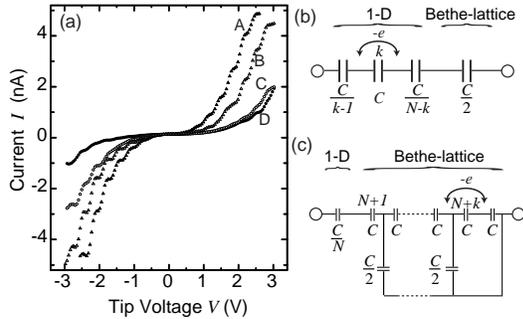}}} \caption{(a) Experimental I-V curves for a 1
$\mu$m-thick Co$_{36}$Al$_{22}$O$_{42}$ film at room temperature.
A, B, C and D refer to different distance of the STM tip from the surface
 of the sample.
The lateral tip position for A and B is different from that 
 for C and D. (b)The equivalent network for the electron tunneling in
the one-dimensional array, where the Bethe-lattice network is replaced
by its total capacitance $C/2$.  (c)The equivalent network for the
electron tunneling in the Bethe-lattice network.}  \label{fig:thick}
\end{figure}

The experimental I-V curves for a 10 nm-thick
Co$_{36}$Al$_{22}$O$_{42}$ film are plotted in Fig. \ref{fig:thin} (a).
There exist two or three granules between the tip and base electrode.
Even at room temperature, the tunneling
current shows a clear CS with a single period.
The I-V curves for a 1 $\mu$m-thick Co$_{36}$Al$_{22}$O$_{42}$ film are
plotted in Fig. \ref{fig:thick}(a).  We also find the CS with a single
period.  Note that for this thick film, 200 to 300 Co granules exist
in the direction perpendicular to the film plane between the tip and the 
substrate.  Let us now analyze our experimental results by using the
theory presented above.

%%%%%%%%%%%%%%%%%%%%%%%%%%%%%%%%%%%%%%%%
%%  figure 3 : I-V triple junction    %%
%%%%%%%%%%%%%%%%%%%%%%%%%%%%%%%%%%%%%%%%

We first examine the I-V curves for the 10 nm-thick film.
A triple tunnel junction model with a bottleneck between the tip and a
granule on the surface is used for the calculation. 
The calculated I-V
curves are shown in Fig. \ref{fig:thin} (b).
Parameter values were chosen in the ranges estimated from
the experiments.  We find that the theoretical curves explain the
experimental ones very well.

%
%
%
%============================================================
%  Thick sample experiments
%============================================================
For the thick film, on the other hand, the conducting paths are considered
to form a
three-dimensional network inside the film; this is a more
complicated system compared to the thin granular films for which the CS was
observed so far\cite{bar-sadeh94,bar-sadeh95,desmicht98}.
The value of the threshold voltage $V_{T}\simeq 0.5$V is not explained
by the one-dimensional array model, because the threshold voltage
$V_{T}$ is proportional to the layer thickness in this model and is
about 100 times larger than that for 10
nm-thick film; this is in contrast with the experimental data.

%
%
%
%============================================================
%  Theory for thick sample
%============================================================
As mentioned before, this discrepancy can be resolved by considering a
network of the conducting paths.  We describe the thick film as a
one-dimensional array connected to a Bethe-lattice network as shown in
Fig. \ref{fig:cartoons}(d).  The bottleneck is
created between the tip and a granule on the surface.  
Therefore, the stable states are
given by $(n_{1},0,\dots,0)$ and the tunneling current jumps when the
number of accumulated electrons $n_{1}$ changes.  The voltage where the
current jumps is given by Eq. (\ref{eq:period}) and the CS
has a single period of $e/C_{1}$ even for a thick film.  The
equivalent network to obtain the threshold voltage for $k$-th junction
$V_{T}^{k}$ is shown in Figs. \ref{fig:thick} (b) and
\ref{fig:thick}(c), where we assume, for simplicity, the same capacitance
$C$ for all junctions.
The key point is that for electron tunneling in the one-dimensional array, 
the Bethe-lattice network is replaced by its total capacitance
as shown in Fig. \ref{fig:thick} (b).   The total capacitance for the
Bethe-lattice is $C/2$ and the bias voltage $V$ at which the tunneling
current jumps (see Eq. \ref{eq:period}) is given by
$V=(e/2C)(N+1)+(e/C)(n_{1}-1)$.  The
experimental results shown in Fig. \ref{fig:thick}(a) are consistent with 
our model with $N\simeq 2$.

%
%
%
%============================================================
%  TMR
%============================================================
The TMR in tunnel junctions with ferromagnetic electrodes
\cite{julliere75,maekawa82} and magnetic granular films \cite{mitani98}
is another attractive topic.  
Recently, the TMR oscillations in
asymmetric double tunnel junctions with ferromagnetic electrodes have
been studied\cite{barnas98,majumdar98}.  The condition for TMR
oscillations is that the tunneling current shows the CS and the
magnetic field dependence of the tunnel resistance is not the same for
all tunnel junctions.  
The TMR oscillations with the same period as the CS may be observed for
magnetic granular films. Our calculation shows that the
amplitude of the TMR oscillation is about 5\% at room temperature for
the curve A in Fig. \ref{fig:thin}(b).

%
%
%
%============================================================
%  Summary
%============================================================

In summary, the electron transport through an array of
tunnel junctions consisting of an STM tip and a granular film has been
studied both theoretically and experimentally.  When the tunnel
resistance between the tip and a granule on the surface is much larger
than those between granules, a bottleneck is created in the array and a
CS with a single period is observed
in the I-V curve.
We predicted that the period of the CS is given by
the capacitance at the bottleneck even in a
thick film with many granules between the tip and the base electrode.
Our STM experiments on 10 nm- and 1$\mu$m-thick Co-Al-O granular
films confirmed the CS with a single period at room temperature. 
TMR oscillations for magnetic granular films are also
predicted.
Our theoretical and experimental studies
provide a new direction for single-electron-spin-electronic devices at
room temperature.
%
%
%============================================================
%  Acknowledgment
%============================================================

We thank P. M. Levy for reading the manuscript. 
This work is supported by a Grant-in-Aid from Scientific Research
Priority Area for Ministry of Education, Science, Sports and Culture
of Japan, a Grant from the Japan Society for Promotion of Science, and
NEDO Japan.
 
%
%
%============================================================
%  References
%============================================================


\begin{thebibliography}{10}
\bibitem{double}
For a double junction system, our classification of the CS is consistent with
  the results in Ref. \cite{hanna91}.
\end{thebibliography}
\end{document}